
%
%
\def\half {{1 \over 2}}
\def\dz{{\partial _z}}

\def\plb{{+\bar l}}

\def\sl{{\Sigma_{l=1}^4}}
\def\sk{{\Sigma_{k=1}^4}}
\def\hp{{h^+}}
\def\hm{{h^-}}
\def\ml{{-l}}

\def\vep {{\varepsilon^+}}
\def\vem {{\varepsilon^-}}
\def\ve {{\varepsilon}}
\def\sm {{\psi^-}}
\def\sp {{\psi^+}}
\def\dzxp {{\dz x^{9+0} +\half\sp\dz\sm+\half\sm\dz\sp}}

\def\xml {{x^\ml}}
\def\Gplb {{\Gamma^\plb}}
\def\Gml {{\Gamma^\ml}}

\tolerance=5000
\footline={\ifnum\pageno>1 \hfil {\rm \folio} \hfil \else \hfil \fi}

\overfullrule=0pt 
\baselineskip=18pt
\raggedbottom
\centerline{\bf The Ten-Dimensional Green-Schwarz Superstring is a}
\centerline{\bf Twisted Neveu-Schwarz-Ramond String}
\vskip 24pt
\centerline{Nathan Berkovits}
\vskip 24pt
\centerline{Math Dept., King's College, Strand,
London, WC2R 2LS, United Kingdom}
\vskip 12pt
\centerline{e-mail: udah101@oak.cc.kcl.ac.uk}
\vskip 12pt
\centerline {KCL-TH-93-12}
\vskip 12pt
\centerline {August 1993}
\vskip 96pt
\centerline {\bf Abstract}
\vskip 12pt

An action for the ten-dimensional Green-Schwarz superstring with
N=2 worldsheet superconformal invariance has recently been used
to calculate superstring scattering amplitudes and prove their finiteness.
In this paper, it is shown that the N=2 stress-energy tensor
for this Green-Schwarz action can be constructed out of the
stress-energy tensor and ghosts of the Neveu-Schwarz-Ramond action by the
standard twisting procedure. In other words, a field redefinition is found
from the GS matter fields into the NSR matter and ghost fields
which transforms the matter part of the two fermionic GS superconformal
generators into the $b$ ghost and shifted BRST current of the
NSR string. In light-cone gauge, this field redefinition reduces
to the usual one relating the light-cone GS and NSR fields.

Although this proves the equivalence of physical vertex
operators in the two superstring formalisms, multiloop amplitudes are
easier to calculate using the Green-Schwarz formalism since manifest spacetime
supersymmetry eliminates the need for spin cuts, GSO projections, and
cutoffs in moduli space.
\vfil
\eject
\vskip 12pt

Soon after the discovery that the GSO-projected Neveu-Schwarz-Ramond
string is spacetime supersymmetric, Green and Schwarz developed
a light-cone gauge superstring formalism which makes eight of these
spacetime supersymmetries manifest.$^1$ Unfortunately, the requirement
of light-cone gauge-fixing complicated the amplitude calculations
due to the presence of non-trivial
interaction-point operators, and only tree and
one-loop four-point superstring amplitudes were explicitly evaluated
using this light-cone Green-Schwarz method. Nevertheless, it was
shown that the light-cone GS fields are related to the light-cone NSR
fields by a field redefinition which transforms the light-cone GS action
and interaction-point
operators into the light-cone NSR action and interaction-point operators,
thereby proving the equvalence of the two light-cone superstring
formalisms.$^{1,2,3}$

In 1986, Friedan, Martinec, and Shenker developed for the NSR formalism
a conformally invariant method for calculating scattering ampltudes
which avoids the problems of light-cone gauge.$^4$ This method involves
evaluating correlation functions of vertex operators on N=1
super-Riemann surfaces, integrating over the N=1 super-moduli, and summing
over spin structures on the surface. Unlike the
conformally invariant method for the bosonic string, the method of
Friedan, Martinec, and Shenker requires vertex operators which depend in
a crucial way on ghost fields.

Although there have been many suggestions for conformally invariant actions
of the Green-Schwarz superstring, only one such action has been
successfully quantized and used to calculate multiloop scattering
amplitudes.$^{5-8}$
This action is actually N=2 superconformally invariant, and amplitudes
are calculated by evaluating correlation functions of vertex operators
on N=2 surfaces, and integrating over the N=2 super-moduli.
Unlike vertex operators in the N=1 superconformally invariant NSR formalism,
the N=2 superconformally invariant GS vertex operators do not
require ghost fields.$^7$

Since after gauge-fixing the superconformal invariance to light-cone gauge,
the NSR and GS light-cone fields are related by a
field redefinition, it is reasonable to ask if the non-gauge-fixed fields
in the two formalisms can also be related to each other. In this paper,
it will be shown that such a field redefinition exists, and not
surprisingly, it transforms matter fields of the conformally invariant
GS formalism into both matter and ghost fields of the conformally
invariant NSR formalism.

Furthermore, it will be shown that this field redefinition
transforms the N=2 stress-energy tensor for the Green-Schwarz matter
fields into a twisted N=2 tensor constructed from a combination of NSR
ghosts and a shifted BRST current$^9$ (the equivalence of these N=2 tensors
was first suggested by Cumrun Vafa$^{10}$).
Shifting the NSR BRST current, $q_{NSR}$,
is necessary to make non-singular the operator product of $q_{NSR}$
with itself, however unlike in the critical
bosonic string, this can be accomplished
without breaking Lorentz invariance by shifting $q_{NSR}\to
q_{NSR}
+\dz(c\xi\eta)+\partial_z^2 c$ where $(c,b,\eta e^\phi , \dz\xi  e^{-\phi})$
are the NSR ghosts.

It should be stressed that although the field redefinition can be used to
relate physical vertex operators in the conformally invariant NSR
and GS formalisms, amplitude calculations appear completely
different using the two different methods$^6$ (this is not surprising since
the integrand of an NSR scattering amplitude is only expected to agree with
the integrand of a GS amplitude up to a total derivative in the bosonic
moduli). The advantage of the NSR formalism is that Lorentz transformations
act linearly on the free fields, while the advantage of the GS formalism
is that some of the spacetime supersymmetries act linearly on the free
fields. So for analyzing properties of superstring amplitudes that are
affected by spacetime supersymmetry, the GS method is usually more
convenient since there is no need to introduce spin cuts or GSO projections.
For example, proving the finiteness of multiloop superstring
amplitudes is much simpler
using the Green-Schwarz formalism$^{8}$ where there is no need to introduce
a cutoff before summing over spin structures,$^{11}$ so
moduli space can be
compactified without ``multiloop ambiguities''.$^{12}$

As discussed in references 6-8, the matter fields of the conformally
invariant Green-Schwarz formalism consist of ten real bosons,
$x^\mu$ ($\mu=0$ to 9), four pairs of right-moving spin $\half$ fermions,
$\Gml$ and $\Gplb$ ($l=1$ to 4), a pair of
right-moving bosons with screening charge $-1$,
$h^+$ and $h^-$, and two pairs of right-moving spin 0 and spin 1
fermions, $\psi^\pm$ and $\ve^\mp$, all with the usual free-field operator
products. For the non-heterotic superstring, the remaining left-moving
fields are the complex conjugates of the right-moving fields, while for
the heterotic superstring, they are the same as in the heterotic sector
of the NSR string.

The N=2 stress-energy tensor for these GS matter fields is:$^{6-8}$
$$J=\Gplb\Gml -\dz\hp +\dz\hm,\quad
G^-=\dz x^\plb \Gml +(\vep+\half\psi^+ \dz x^{9-0}) e^{-\hp},\quad
G^+=\dz\xml\Gplb+ \eqno(1)$$
$$(\vem+\half \dz x^{9-0}\psi^- )
((\dzxp)
 e^\hp +e^{-\hm})-e^\hp ((\dz h^+ +\dz h^-)\dz\psi^- +{3\over 4}
\partial^2_z \psi^-),$$
$$L=
\dz x_\mu \dz x^\mu -\half (\Gplb\dz\Gml +\Gml\dz\Gplb)-\vep
\dz\psi^- -\vem\dz\psi^+ +\dz \hp\dz\hm +\half(\partial^2_z \hp
+\partial^2_z \hm ),$$
where $x^{9+0}\equiv x^9\pm x^0$, $x^\ml\equiv
x^l-ix^{l+4}$, and $x^\plb\equiv x^l+ix^{l+4}$.

The first step in relating these GS fields to NSR fields is to
redefine $x^\mu$ in such a way that it commutes with $G^-$. This is
done by defining
$x_{new}^{9+0}\equiv x_{old}^{9+0} +\half\psi^+\psi^-$ and
$x_{new}^\ml\equiv x_{old}^\ml-e^\hp \psi^-\Gml$ ($x_{new}^\mu$ was
called $x_+^\mu$ in reference 8). In order to preserve the free-field
operator product relations, one must also redefine all other GS fields,
$\Phi$, using
$$\Phi_{new}\equiv e^{\int dz (\half\dz x^{9-0}\psi^+\psi^- -
\dz x^\plb e^\hp \psi^- \Gml)}~\Phi_{old}~
e^{-\int dz (\half\dz x^{9-0}\psi^+\psi^- -
\dz x^\plb e^\hp \psi^- \Gml)}.\eqno(2)$$
In terms of these newly defined GS fields, the N=2 stress-energy tensor
of equation (1) takes the following form:
$$J=\Gplb\Gml -\dz\hp +\dz\hm,\quad
G^-=\vep e^{-\hp}, \eqno(3)$$
$$G^+=\dz\xml\Gplb+
+\dz x^\plb \vem\psi^-\Gml e^{\hp-\hm} +\dz x^{9+0}\vem e^\hp
+\dz x^{9-0} \psi^- e^{-\hm}$$
$$+\psi^- e^\hp
(\dz x_\mu \dz x^\mu -\Gml\dz\Gplb
-\vem\dz\psi^+ +\dz \hp\dz\hm +\partial^2_z \hm)
+\vem e^{-\hm}+\dz[\psi^- e^\hp (\Gml\Gplb+\dz \hp-\dz\hm)]$$
$$
+{3\over 2} e^\hp\partial^2_z \psi^-+2e^\hp\psi^-\partial^2_z \hp
+2(\dz \hp)^2 e^\hp\psi^- +3e^\hp\dz\hp\dz\psi^-,$$
$$L=
\dz x_\mu \dz x^\mu -\half (\Gplb\dz\Gml +\Gml\dz\Gplb)-\vep
\dz\psi^- -\vem\dz\psi^+ +\dz \hp\dz\hm +\half(\partial^2_z \hp
+\partial^2_z \hm ).$$

It is now straightforward to guess the following
field transformation from the new
Green-Schwarz matter fields to the Neveu-Schwarz-Ramond matter and
ghost fields:
$$x^\mu\to x^\mu,\quad
\Gml\to \xi e^{-\phi-\sigma_l},\quad
\Gplb\to \eta e^{\phi+\sigma_l},\eqno(4)$$
$$
\psi^+\to e^{\half(\phi+\sigma_0+\sl \sigma_l)},\quad
\psi^-\to c\xi e^{\half(-3\phi+\sigma_0-\sl \sigma_l)},\quad
\vem\to e^{-\half(\phi+\sigma_0+\sl \sigma_l)},\quad
\vep\to b\eta e^{\half(3\phi-\sigma_0+\sl \sigma_l)},$$
$$e^\hp\to\eta e^{\half(3\phi-\sigma_0+\sl\sigma_l)},\quad
e^{-\hp}\to\xi e^{\half(-3\phi+\sigma_0-\sl\sigma_l)},$$
$$
e^\hm\to c\xi\dz\xi\, e^{\half(-5\phi-\sigma_0-\sl\sigma_l)},\quad
e^{-\hm}\to b\eta\dz\eta\, e^{\half(5\phi+\sigma_0+\sl\sigma_l)},$$
where $e^{\pm \sigma_0}$ and
$e^{\pm \sigma_l}$ are the ten NSR fermionic matter fields and
$(c,b,\eta e^\phi, \dz\xi e^{-\phi})$ are the NSR ghost fields.
Since the eight GS fermionic fields $\Gml e^\hp$ and $\Gplb e^{-\hp}$
transform into $e^{\pm\half(\phi-\sigma_0+\sk\sigma_k)\mp\sigma_l}$, this
field transformation has the expected property that in light-cone gauge
(where $\hp=\phi-\sigma_0=0$), it reproduces the ``triality'' rotation
of an SO(8) spinor into an SO(8) vector.$^2$

Note that after bosonizing all right-moving fields including
$\psi^\pm$, $\varepsilon^\mp$,
$\Gml$, $\Gplb$, $c$, $b$, $\eta$,
and $\xi$, both the GS and NSR variables can
be described by eight chiral bosons
(the lattice
constructed out of these eight bosons is probably
closely related to the lattices
discussed in reference 13).
Since the field transformation of equation (4) acts linearly on these
bosons, it is invertible. However only GSO-projected combinations of
NSR fields (i.e., combinations whose operator product with the
spacetime-supersymmetry
generator contains no square-root cuts) are invariant under
$2\pi$ shifts of the Green-Schwarz
chiral bosons. For example, the NSR ghost field
$\beta=\dz\xi e^{-\phi}$ can not be expressed in terms of single-valued
GS fields (since $\beta$ is one
of the generators of the N=3 superconformal algebra found in reference 9,
only an N=2 subgroup of this algebra is present in the GS superstring).

Under the above field transformation, the N=2 stress-energy tensor of
equation (3) transforms into the following combination of NSR ghosts
and shifted BRST tensor:
$$J\to cb+\eta\xi,\quad G^-\to b,\quad
L\to L_{NSR}+\half\dz(bc+\xi\eta),\eqno(5)$$
$$\quad G^+\to\eta e^\phi G_{NSR} +c L_{NSR}
-c\dz c b +\eta\dz\eta b +\dz(c\xi\eta)+\partial^2_z c ~=q_{NSR}+
\dz(c\xi\eta)+\partial^2_z c.  $$
It is easy to check that this tensor satisfies the usual N=2
operator-product algebra, and in particular, that $G^+$ has a
non-singular operator-product with itself.

To prove that physical GS vertex operators transform into physical
NSR vertex operators, let $V_{GS}(z)$ be a vertex operator constructed
out of GS matter fields such that $\int dz V_{GS}(z)$ is N=2
superconformally invariant (for examples of such vertex operators, see
reference 7). Then $[G^+,\int dz V_{GS}]$=0 implies that $[Q_{NSR},
\int dz V_{NSR}]$=0 where $V_{NSR}$ is the transformed $V_{GS}$.
Furthermore, $[J,\int dz V_{GS}]=0$ implies that the NSR ghost field
$\xi$ only occurs in the combination $(c\xi)$, and since
$[b,\int V_{NSR}]=[G^-,\int V_{GS}]=0$, the zero mode of $\xi$ can not
appear in $V_{NSR}$, so $V_{NSR}$ is physical.

It is not true that all physical NSR vertex operators come from physical
GS vertex operators since $[Q_{NSR},V_{NSR}]=0$ does not imply that
$[b, V_{NSR}]=0$. However if the physical NSR vertex operator is GSO-projected
and is constructed
entirely out of matter fields (e.g., the integrated vertex operator for
an even G-parity
Neveu-Schwarz state of ghost-number zero), it is easy to show that
the corresponding $V_{GS}$ is single-valued
and is N=2 superconformally invariant. Since the spacetime-supersymmetry
generators commute with the N=2 stress-energy tensor,
this is also true for any NSR vertex operator which can be obtained from
a ghost-free physical vertex operator by a supersymmetry transformation.

In reference 6, it was shown that the same GS state can be represented
by vertex operators of different instanton number, where the instanton
charge is defined as $\half\int dz (\vep \psi^--\vem\psi^++\dz\hm-
\dz \hp)$. Since
the instanton charge transforms into $\int dz(\xi\eta-\dz\phi)$ and since
the U(1) charge, $\int dz J$, transforms into $\int dz (cb+
\eta\xi)$, the instanton-number of a U(1)-invariant GS state is precisely
the ghost number of the corresponding NSR state.
Note that
the GS spacetime-supersymmetry generators which were defined in reference 9,
$S^-_{\alpha}$ and $S^+_{\alpha}$,
transform into the NSR spacetime-supersymmetry
generators that lower and raise the ghost number.$^4$ For
example,
$$S^-_{-----}=\vem-\half\dz x^{9-0}\psi^-\to \int dz e^{-\half(\phi+\sigma_0
+\sl\sigma_l)},\eqno(6)$$
$$S^+_{-++++}=\vep-\half\dz x^{9-0}\psi^+\to \int dz [b\eta
e^{\half(3\phi-\sigma_0 +\sl\sigma_l)}-e^{\half(\phi-\sigma_0+\sl\sigma_l)}
(\dz x^{9-0} e^{\sigma_0}+\dz x^{\plb} e^{-\sigma_l})].$$

Because tree-level superstring amplitudes are calculated by taking
operator products of ghost-free vertex operators, there are no essential
differences between the NSR and GS methods at tree-level. For writing
these amplitudes in a manifestly Lorentz-invariant form,
the NSR method has an advantage since the Neveu-Schwarz vertex operators
are manifestly Lorentz covariant, while for making spacetime supersymmetry
manifest, the GS method has an advantage since there are no spin cuts
and since the fields transform linearly under the two supersymmetries
described in equation (6).

For multiloop amplitude calculations, however, there is a major difference
between the two formalisms. Using the NSR method, amplitudes are calculated
by evaluating correlation functions of vertex operators on N=1
super-Riemann surfaces, integrating over the surface moduli, and then
summing over the spin structures.$^4$
Since the amplitude is not divergence-free
until after summing over spin structures, a cutoff must be introduced for the
integration region of the moduli,$^{12}$ which can
only be removed after summing over spin structures.
Although the precise form of this
cutoff can be determined from unitarity considerations,$^{11}$ the resulting
closed-form expression for the NSR amplitude after summing over spin
structures is too complicated for an analysis of finiteness.

In the GS method, multiloop amplitudes are calculated by evaluating
correlation functions of vertex operators on N=2 super-Riemann surfaces,
rather than N=1 super-Riemann surfaces. Because of spectral flow on N=2
surfaces, all GS fields can have integer conformal weights, eliminating the
need for spin structures.$^{14}$ Although integration over the N=2 moduli in
the GS method is considerably different
from in the NSR method,
there is no problem in performing this integration and finding explicit
expressions for GS multiloop scattering amplitudes.$^6$
Unlike in the NSR case, it is straightforward to prove that these
amplitudes are finite.$^8$

In retrospect, it is not surprising that multiloop superstring amplitudes
look very different when calculated using the two different formalisms.
Although in light-cone gauge, the integrands of NSR and GS amplitudes
are expected to coincide; in a conformally invariant gauge, the
integrands are only expected to agree up to a total derivative in
the bosonic moduli of the surface. It is precisely the contribution
of this total derivative which causes complications when using the
NSR method since the presence of a cutoff in moduli space can lead
to surface terms in the scattering amplitude.$^{12}$

\vskip 24pt
\centerline {\bf Acknowledgements}
\vskip 12pt
I would like to thank M. Bershadsky, J. Figueroa-O'Farrill, W. Lerche,
W. Siegel, and C. Vafa
for useful discussions, and the SERC for its financial support.

\vskip 24pt

\centerline{\bf References}
\vskip 12pt

\item{(1)} Schwarz,J., Phys.Rep.89 (1982), p.223.

\item{(2)} Witten,E., `D=10 Superstring Theory' in Fourth Workshop
on Grand Unification, ed. P.Langacker et.al., Berkhauser (1983), p.395.

\item{(3)} Mandelstam,S., Prog.Theor.Phys.Suppl.86 (1986), p.163.

\item{(4)} Friedan,D., Martinec,E., and Shenker,S., Nucl.Phys.B271
(1986), p.93.

\item{(5)} Berkovits,N., Nucl.Phys.B379 (1992), p.96., hep-th no. 9201004.

\item{(6)} Berkovits,N., Nucl.Phys.B395 (1993), p.77, hep-th no. 9208035.

\item{(7)} Berkovits,N., Phys.Lett.B300 (1993), p.53, hep-th no. 9211025.

\item{(8)} Berkovits,N., ``Finiteness and Unitarity of Lorentz-covariant
Green-Schwarz Superstring Amplitudes'', King's College preprint
KCL-TH-93-6, March 1993, to appear in Nucl.Phys.B, hep-th no. 9303122.

\item{(9)} Bershadsky,M., Lerche,W., Nemeschansky,D., and Warner,N.P.,
``Extended N=2 Superconformal Structure of Gravity and W-Gravity Coupled
to Matter'', Harvard preprint HUTP-A061/92, November 1992, hep-th no.
9211040.

\item{(10)} Vafa,C., private communication.

\item{(11)} Mandelstam,S., Phys.Lett.B277 (1992), p.82.

\item{(12)} Atick,J. and Sen,A., Nucl.Phys.B296 (1988), p.157.

\item{(13)} Lerche,W., Schellekens,A., and Warner,N.P., Phys.Rep.177 (1989),
p.1.

\item{(14)} Ooguri,H. and Vafa,C., Nucl.Phys.B361 (1991), p.469.

\end